\documentclass[aps,prl,twocolumn,showpacs,superscriptaddress,groupedaddress]{revtex4-2}  
\usepackage[T1]{fontenc}
\usepackage{lmodern} 
\usepackage{graphicx}  
\usepackage{dcolumn}   
\usepackage{bm}        
\usepackage{amssymb}   
\usepackage{amsmath}
\usepackage{bbold}
\usepackage{array}
\usepackage{makecell}
\usepackage{braket}
\usepackage{physics}
\usepackage{titlesec} 
\usepackage[colorlinks,citecolor=blue,linkcolor=blue,urlcolor=blue,hypertexnames=true]{hyperref}
\usepackage{subfigure}
\usepackage{supertabular,booktabs}

\usepackage[usenames,dvipsnames,svgnames,table]{xcolor} 
\allowdisplaybreaks

\newcommand\superequiv{\mathrel{\rlap{\raisebox{\fontdimen22\textfont2}{$=$}}\raisebox{-0.5\fontdimen22\textfont2}{$ = $}}}

\begin{document}

\widetext


 \title{\textcolor{Sepia}{\textbf{\LARGE  Krylov Complexity in Quantum Field Theory }}}


\author{Kiran Adhikari$^{1}$}

\author{Sayantan Choudhury$^{2,3}$}
\email{\text{corresponding author:}\\ sayantan\_ccsp@sgtuniversity.org,\\  sayanphysicsisi@gmail.com}
\author{Abhishek Roy.$^{4}$}

\affiliation{${}^{1}$RWTH Aachen University, D-52056, Aachen, Germany,}
\affiliation{ ${}^{2}$Centre For Cosmology and Science Popularization (CCSP),\\
        SGT University, Gurugram, Delhi- NCR, Haryana- 122505, India,}
\affiliation{${}^{3}$International Centre for Theoretical Sciences, Tata Institute of Fundamental Research (ICTS-TIFR), Shivakote, Bengaluru 560089, India,}
\affiliation{${}^{4}$Department of Physics, Indian Institute of Technology Jodhpur,
Karwar, Jodhpur 342037, India.}

\begin{abstract}
In this paper, we study the Krylov complexity in quantum field theory and make a connection with the holographic "Complexity equals Volume" conjecture. When Krylov basis matches with Fock basis, for several interesting settings, we observe that the Krylov complexity equals the average particle number showing that complexity scales with volume. Using similar formalism, we compute the Krylov complexity for free scalar field theory and find surprising similarities with holography. We also extend this framework for field theory where an inverted oscillator appears naturally and explore its chaotic behavior.

\end{abstract}

\maketitle

\section{{\textbf{ \large Introduction}}} \label{sec:introduction}
Recently, there has been a huge interest of applying quantum information ideas into high energy physics and cosmology \cite{Chapman:2021jbh,Chapman:2021eyy,Faulkner:2022mlp,Shaghoulian:2022fop,Adhikari:2021ckk}. One method that is getting the attention is complexity \cite{Brown:2017jil,Susskind:2014moa,Brown:2015lvg,Maldacena:2015waa} which characterizes the difficulty of preparing a certain quantum state or applying a certain unitary operator. Indeed, there are various measures of complexity in literature \cite{Nielsen1,Nielsen2,Nielsen3,Nielsen4,Balasubramanian:2022tpr}. Krylov/K complexity is one such tool 
 \cite{Caputa:2021sib,Parker:2018yvk,Roberts:2018mnp,Rabinovici:2020ryf,Barbon:2019wsy,Jian:2020qpp,Dymarsky:2019elm,Dymarsky:2021bjq, Hornedal:2022pkc} where one attempts to study the Heisenberg evolution of certain initial operator. The evolution can become extremely complicated depending on the hamiltonian and initial operator, and this is captured by the Krylov complexity. This new measure of complexity has now been explored from black holes to conformal field theories \cite{Roberts:2018mnp, Jian:2020qpp,Dymarsky:2021bjq,Caputa:2021ori}.

The primary motivation to study complexity in Quantum Field Theory(QFT)  arises from various holographic conjectures. Indeed, it was  conjectured that complexity can explore beyond black hole horizons via “complexity = volume” and “complexity = action” conjectures \cite{Susskind:2014moa, Brown:2015lvg}. After this work in \cite{Susskind:2014moa, Brown:2015lvg} , circuit complexity using Nielsen's method \cite{Nielsen1,Nielsen2,Nielsen3,Nielsen4} has been computed in quantum field theory and even in cosmology \cite{Balasubramanian:2021mxo, Balasubramanian:2019wgd,Adhikari:2021ckk, Choudhury:2021brg, Adhikari:2021ked, Adhikari:2021pvv,Choudhury:2020hil,Bhargava:2020fhl, Bhattacharyya:2020art, Adhikari:2022oxr, Chapman:2021jbh,Jefferson:2017sdb,Bhattacharyya:2020rpy,Chapman_2018}. However, the issue with using Nielsen's measure for high energy physics computations is that it is very ambiguous depending on the various factors such as choices of gates, reference and target gates, and tolerance. However, Krylov's complexity is free from such ambiguities, making it an ideal choice for holographic and QFT settings. 

In this paper, we study Krylov complexity for quantum field theories and make a comparison with holography. First we give a review on Krylov complexity and show that it also satisfies properties of Nielsen's measure like continuity and triangle inequality which makes it attractive for high energy physics.  We then show that when Krylov basis matches with fock basis, Krylov complexity equals the average particle number. Since volume is proportional to particle number, this provides hints towards the relevance of Krylov complexity with volume. Then, we compute the Krylov complexity for free scalar field theory following the philosophy of \cite{Jefferson:2017sdb,Chapman_2018}. Having done this, we make an comparison with holography and previous works, and find that even for such simple settings, it shows signs of qualitative matches. Krylov complexity is often studied to explore the chaotic behavior of the system. However, our primary goal in this paper is to study the complexity for QFT and holography, therefore full study of chaotic behavior is beyond our scope. However, we provide a field theory where inverted oscillator appears naturally, and we comment on it's chaotic property.

\section{{ \large Krylov complexity}}
\label{sec:E2}

 In literature, one can find a zoo of complexity measure among which Nielsen's measure \cite{Nielsen1, Nielsen2, Nielsen3, Nielsen4} is getting traction in the high energy physics community particularly because of it's continuity nature \cite{Jefferson:2017sdb,Chapman_2018} . This new interest in Krylov complexity comes from the various ambiguities in Nielsen's measure such as arbitrary choice of gates, reference and target states and tolerance which makes it difficult to properly define in QFTs or holography. Fortunately, Krylov complexity is free of such choices and also satisfies the desired property of Nielsen's measure, like continuity and triangle inequality. For a detailed overview of Krylov complexity, refer \cite{Caputa:2021sib,Bhattacharjee:2022vlt, Balasubramanian:2022tpr, Hornedal:2022pkc}.

Let us consider a Hamiltonian $H$ and time-dependent Heisenberg operator $\mathcal{W}(t)$. Heisenberg equations describes the evolution of the operator as
\begin{equation}
    \partial_t\mathcal{W}(t) = i[H,\mathcal{W}(t)]
\end{equation}
whose solution is given by
\begin{equation}
    \mathcal{W}(t)  =  e^{iHt}\mathcal{W}e^{-iHt}
\end{equation}
where, $\mathcal{W}=\mathcal{W}(0)$. Defining Liouvillian super-operator $\mathcal{L}_X$ as $\mathcal{L}_X Y = [X,Y]$, time evolution of $\mathcal{W}(t) $ can be expressed as
\begin{equation}
\begin{aligned}
\label{eq:hopping}
 \mathcal{W}(t)  &=  \sum_{n=0}^\infty \frac{(it)^n}{n!}\mathcal{L}_H^n \mathcal{W} 
\end{aligned}
\end{equation}
As time evolves, initial operator spreads and more complicated nested commutators need to be taken into consideration. Krylov complexity measures this complexity in a precise manner. 
Representing the action of $\mathcal{L}$ as $\tilde{\mathcal{W}}_n =\mathcal{L}^n \mathcal{W}$, $\mathcal{W}(t) $ becomes
\begin{equation}
\label{eq:schroseries}
    \mathcal{W}(t) = e^{i\mathcal{L}t}\mathcal{W} = \sum_{n=0}^\infty \frac{(it)^n}{n!}\mathcal{L}^n \mathcal{W} = \sum_{n=0}^\infty \frac{(it)^n}{n!}\tilde{\mathcal{W}}_n
\end{equation}
Eq. (\ref{eq:schroseries})  can be interpreted as a Schrödinger's equation where $\mathcal{W}(t)$ are "operator's wave functions", $\mathcal{L}$ is the Hamiltonian and "Hilbert space vectors" are
\begin{equation}
    \mathcal{W} \superequiv |\mathcal{\tilde{W}}), \mathcal{L}^1\mathcal{W} \superequiv |\mathcal{\tilde{W}}_1), \mathcal{L}^2\mathcal{W} \superequiv |\mathcal{\tilde{W}}_2),
    \mathcal{L}^3\mathcal{W} \superequiv |\mathcal{\tilde{W}}_3),\dots
\end{equation}
There is no  guarantee that $|\mathcal{\tilde{W}}_n)$ form an orthonormal basis a prior but with Lanczos algorithm, we can construct orthonormal Krylov basis 
$|\mathcal{W}_n)$. We have described the Lancoz algorithm in \ref{sec:lancoz}. Having obtained the Krylov basis, the time evolved operator $\mathcal{W}(t)$ is
\begin{equation}
\label{eq:schrodinger}
    |\mathcal{W}(t)) = e^{i\mathcal{L}t}|\mathcal{W}) = \sum_n i^n \phi_n(t)  |\mathcal{W}_n)
\end{equation}
where $\phi_n(t)$ are real amplitudes, and $|\phi_n|^2$ sums up to one. Krylov complexity/K-complexity can now be written as
\begin{equation}
    K (t) = \sum_n n |\phi_n|^2
\end{equation}
It is also possible to obtain a physical intuition of Krylov complexity \cite{Dymarsky:2021bjq,Dymarsky:2019elm}. We can think of particle hopping on a one dimensional chain of (\ref{eq:hopping}), and Krylov complexity then looks like an average position on the chain. 

One reason why Nielsen's measure of complexity is mostly used in high energy physics is because of it's continuity nature. Nielsen's complexity enjoys some desirable features such as continuity, and triangle inequality. Here, we would like to point that Krylov complexity also has the similar features. Since operator wave function $\phi_n(t)$ are the solutions of Schrödinger like equation, $\phi_n \in C^0$ and then $K(t) \in C^0$. 

In order to satisfy the triangle inequality, for the operator $W = 0_1 0_2$, the Krylov complexity should satisfy $K_W(t) \leq K_{0_1}(t)K_{0_2}(t)$. To see this, let us imagine that $0_1$ spreads on the x-axis, while $K_{0_2}(t)$ spreads on the y-axis. $ K_{0_1}(t)$ measures the average position on the x-axis, while $K_{0_2}(t)$ on the y-axis. Now, $K_W(t)$ measures the average position on this two-dimensional plane which is triangle like bounded by the walk on x and y line. This idea can be exploited from \cite{Dymarsky:2021bjq,Dymarsky:2019elm}

\section{{ \large Krylov complexity and Volume}}
\label{sec:KeqV}
In \cite{Hornedal:2022pkc, Caputa:2021sib} showed that for the generalized coherent states evolving according to the general displacement operator $D(\xi) = e^{\xi L_+ - \xi L_ -}$, where $\xi = it$ saturates the complexity bound, and therefore are some of the most interesting systems to explore. A overall review of generalized coherent states can be found in \cite{Perelomov:1986tf}. Now, we will show that for these systems, Krylov complexity has surprising relationship with the volume. 

The volume $V$ is proportional to the number of the particles $N$, $V \approx N$. For example, for a spatial volume $V$ with $d-1$ spatial dimension where $n$ particles are placed in each spatial direction with $\delta$ spacing, volume is simply given as $V = n^{d-1} \delta^{d-1} = N\delta^{d-1}$. Therefore, if we can somehow relate the Krylov commplexity to the number statistics, then we could also relate it to the volume. Our approach is as follows. QFT and many body quantum systems are often formulated in the concept of occupation number representation/Fock space. The occupation number $n$ indicates the particle number of the considered state. For generalized coherent states, fock states are basically the Krylov basis. Therefore, in the fock basis particle number probability distribution $P(n)$ is essentially the same as squared of the operator amplitude $\phi(n)^2$ of the operator (\ref{eq:schrodinger}). Given the probability distribution $P(n)$, the average particle number, $\bar{n}$ can be calculated which is then same as Krylov complexity:
\begin{equation}
    \bar{n} = \sum_n n P(n) = \sum_n n\phi(n)^2 = K(t)
\end{equation}

As a simple example, let us consider a two mode squeezed vacuum states which often comes in the context of quantum optics and even in QFT, black hole physics and  cMERA circuit too \cite{Chapman_2018,Haegeman:2011uy, Nozaki:2012zj}.
 \begin{align}
     \ket{SQ(k,\tau)} &= \hat{S}_k(r_k,\phi_k) \ket{0_k,0_{-k}} \nonumber\\
     &= \frac{1}{\cosh r_k}\sum_{n= 0}^\infty e^{-2in\phi_k}\tanh^n r_k\ket{n_k,n_{-k}}
 \end{align}
 where $r_k$ and $\phi_K$ are squeezing parameters, $\ket{n_k,n_{-k}}$ fock states and $\ket{0,0}$ are vacuum states. For two mode squeezing, Krylov basis is the standard two oscillator fock space $|\mathcal{O}_{n} )  =    \ket{n_k,n_{-k}}$. The squared of the operator wave function is the same as the particle number probability distribution
\begin{equation}
     \phi_n^2 = P(n) = \frac{\tanh^{2n} r_k}{\cosh^2 r_k}
\end{equation}
Then, the Krylov complexity equals the average particle number as
\begin{equation}
    K = \sum_n  n |\phi_n|^2 = \sinh^2r_k = \sum_n n P(n) = \bar{n}
\end{equation}
In \cite{Hornedal:2022pkc, Caputa:2021sib}, three main families of generalized coherent states are considered. The first is based on SL(2,R) algebra so called Perelomov coherent states. Just like for two mode squeezed states, we can expand the perelomov coherent states on the fock basis, and obtain the particle number probability distribution. Since fock basis are Krylov basis for these states, average particle number for the Perelomov coherent states equals to the Krylov complexity. Infact, two mode squeezed states also belong to the same SL(2,R) algebra. The second family is standard coherent states belonging to the Heisenberg-Weyl algebra
\begin{equation}
    \ket{z = i \alpha t} = e^{-|z|^2/2} \sum_{n=0}^\infty \frac{z^n}{\sqrt{n!}} \ket{n}
\end{equation}
where $\ket{z}$ are coherent states and $\ket{n}$ are fock states. Again, the particle number probability distribution of the coherent state equals the squared of the operator wave function, $e^{-|z|^2}z^{2n}/n!$. Therefore, both Krylov complexity and average particle number is $z^2 = \alpha^2 t^2$. The third family is spin coherent states belonging to SU(2) algebra, and in the limit of high spins, it becomes the same as standard coherent states via contraction process. So, the similar argument can be applied here too. Finally, we will mention the Krylov complexity for the standard and inverted harmonic oscillators belonging to SL(2,R) algebra of Perelomov coherent states \cite{Balasubramanian:2022tpr, Hornedal:2022pkc, Caputa:2021sib}
\begin{equation}
  K_{\text{HO}}(t) =  \frac{\sin^2(\omega_k t/2)}{\sinh^2(\beta \omega_k /2)}, K_{\text{IHO}}(t) =  \frac{\sinh^2(\omega_k t/2)}{\sin^2(\beta \omega_k /2)}
\end{equation}
where $\omega$ is the oscillator frequency. This computation follows from the generalisation of the notion of Krylov Complexity to the Spread Complexity as presented in \cite{Balasubramanian:2022tpr} These complexity measure will become relevant while computing complexity for QFT in later sections. Relation with volume for these systems follows the same as for SL(2,R) algebra. Krylov complexity for other relevant examples like SYK and Thermofield double has also been computed \cite{Caputa:2021sib,Parker:2018yvk,Roberts:2018mnp,Rabinovici:2020ryf,Barbon:2019wsy,Jian:2020qpp,Dymarsky:2019elm,Dymarsky:2021bjq, Hornedal:2022pkc}.

\section{ \large Krylov complexity in free scalar quantum field theory}
\label{sec:qft}
Having reviewed the concept of Krylov complexity, we will now apply it in the context of free scalar quantum field theory. For this, we will follow the pioneering works done in \cite{Jefferson:2017sdb,Chapman_2018} where they computed complexity for free bosonic quantum field theories in $d$ space-time dimensions using Nielsen's geometric method. The work done in \cite{Chapman_2018} is on momentum representation while \cite{Jefferson:2017sdb} is on space representation. For our work, we will mainly follow the notation and philosophy in \cite{Jefferson:2017sdb} but our methods can be used for momentum space technique of \cite{Chapman_2018} too. In \cite{Jefferson:2017sdb}, the complexity for the quantum field theory was obtained by first placing the theory on the $d-1$ periodic lattice with lattice spacing $\delta$ and Length $L$ in all directions which regularize the divergences. Then, the theory has the structure of $N^{d-1} = (L\ \delta)^{d-1}$ coupled harmonic oscillators. Therefore, in order to compute the Krylov complexity for quantum field theories, we can compute the Krylov complexity for two coupled harmonic oscillators and generalize it to $N^{d-1}$ oscillators. Readers familiar with computing entanglement entropy in quantum field theories can see the similarities with this method. 

\begin{figure*}[htb!]
	\centering
	\includegraphics[width= 0.65 \textwidth]{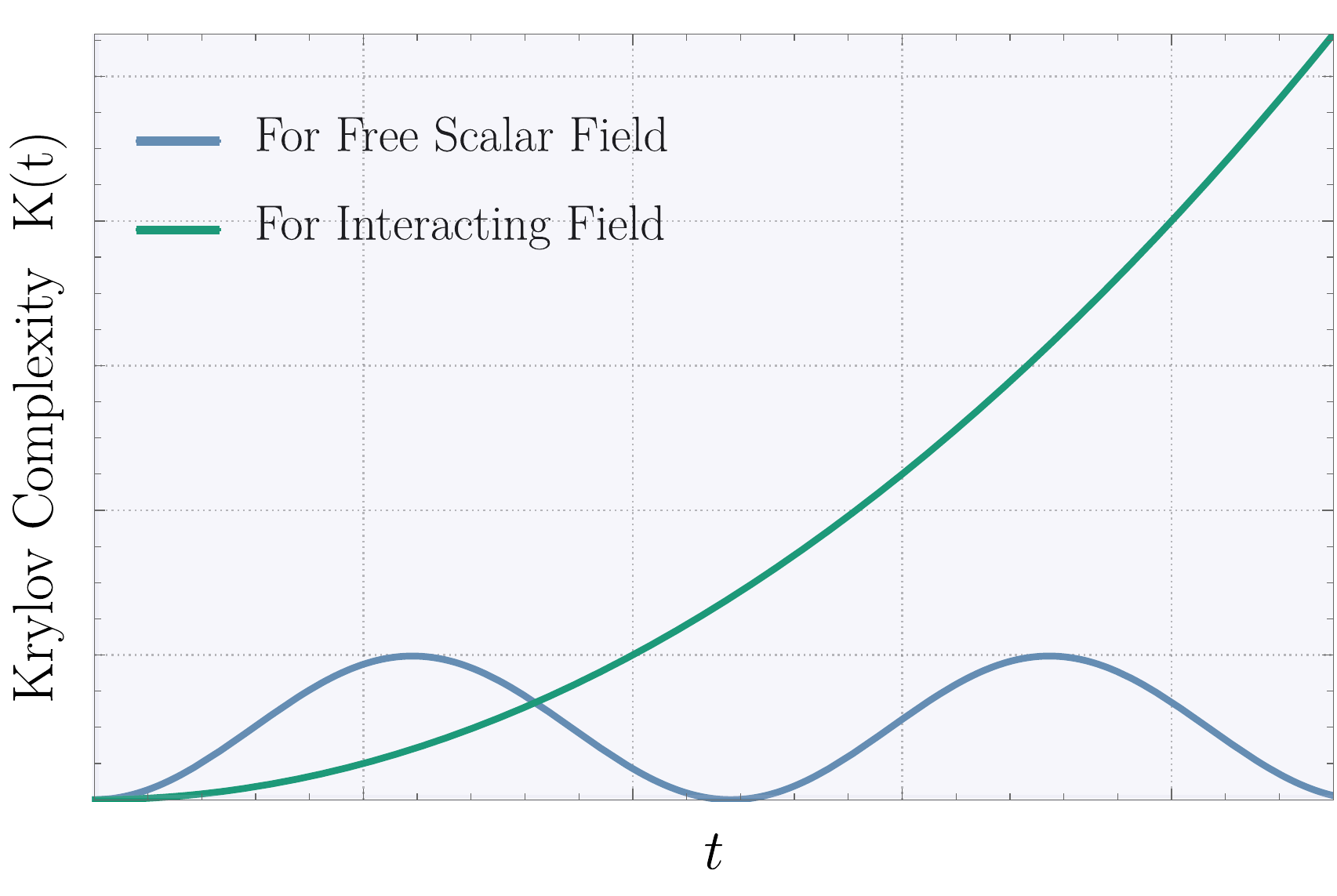}
	\caption{Behavior of Krylov complexity $K(t)$ as a function of time $t$ for free and chaotic interacting field theory. Since complexity for chaotic theory grows exponentially in time (signature of chaos), we have amplified the complexity for free theory in order to have a  better comparison. Without amplification, complexity for free theory is bounded by chaotic theory.}
	\label{fig:KrylovQFT}
\end{figure*}

The Hamiltonian for the massive free scalar field theory in $d-1$ spacial dimension is given by:
\begin{equation}
\label{eq:roHami}
    \text{H} = \frac{1}{2} \int d^{d-1}\left[  \pi(x)^2 + \Vec{\nabla} \phi(x)^2 + m^2\phi(x)^2 \right]
\end{equation}
After discretizing the theory in square lattices, the theory can be reduced to a bunch of coupled harmonic oscillators. As mentioned before, we will start with the two coupled harmonic oscillators and generalize it further. The two coupled harmonic oscillators, in the normal basis the Hamiltonian takes the form (refer appendix \ref{qftlattice} for details.)

\begin{equation}
\label{eq:decoupled}
\text{H}=\textstyle \frac{1}{2}\Big[  \tilde{p}_0^2 + \tilde{p}_1^2 + \tilde{\omega}_0^2\tilde{x}_0^2+ \tilde{\omega}_1^2\tilde{x}_1^2  \Big] .
\end{equation}
which is nothing but two decoupled simple harmonic oscillators. In section (\ref{sec:KeqV}), we gave the expression of Krylov complexity for several interesting examples. One of them was simple harmonic oscillator whose Krylov complexity is given by 
\begin{equation}
  K_{\text{HO}}(t) =  \frac{\sin^2(\omega_k t/2)}{\sinh^2(\beta \omega_k /2)}
\end{equation}
Using the triangle inequality, Krylov complexity for the hamiltonian (\ref{eq:decoupled}) can be expressed as a  sum of Krylov complexity for the individual harmonic oscillator
\begin{equation}
    K(t) = \sum_{k \in [0,1]} \frac{\sin^2(\tilde{\omega}_k t/2)}{\sinh^2(\beta \tilde{\omega}_k /2)}
\end{equation}
Having done this, we can now generalize it to an arbitrary number of coupled oscillators. For simplicity, we will begin with $N$ oscillators on a one-dimensional circular lattice i.e. on $d= 1+1$ space-time dimension whose Hamiltonian is given by:
\begin{equation}
    H = \frac{1}{2} \sum_{a=0}^{N-1} \left[ p_a^2 + \omega^2x_a^2  + \omega^2(x_a- x_{a+1})^2 \right]
\end{equation}
where, periodic boundary condition implies $x_{a+N} =x_a $ and we have set $M_a = 1$. The Hamiltonian then becomes decoupled as explained in \ref{qftlattice}, 
\begin{equation}
    H = \frac{1}{2} \sum_{k=0}^{N-1} \left[  |\tilde{p}_k|^2 + \tilde{\omega}_k^2 |\tilde{x}_k|^2 \right]
\end{equation}
where, $\tilde{\omega}_k^2 = \omega^2 + 4\Omega^2 \sin^2 \frac{\pi k }{N} $.  Here $\Omega$ physically represents the inter-mass coupling parameter which signifies interaction in the present context of discussion.  The Krylov complexity for (\ref{eq:h2}) is then given by
\begin{equation}
  K(t) =   \sum_{k=0}^{N-1}\frac{\sin^2(\tilde{\omega}_k t/2)}{\sinh^2(\beta \tilde{\omega}_k /2)}
\end{equation}
We can then generalize to arbitrary dimensions as
\begin{equation}
\label{eq:KrylovQFT}
   K(t) =   \sum_{\vec{k}=0}^{N-1}\frac{\sin^2(\tilde{\omega}_{\Vec{k}} t/2)}{\sinh^2(\beta \tilde{\omega}_{\Vec{k}} /2)}  
\end{equation}
where $k_i$ are the components of the momentum vector $\Vec{k} = (k_1, k_2, ... , k_{d-1})$, and the normal-mode frequencies are given by:
\begin{equation}
 \tilde{\omega}_{\Vec{k}}^2 = m^2 + \frac{4}{\delta^2} \sum_{i=1}^{d-1}\sin^2\frac{\pi k_i}{N}  
\end{equation}
In figure \ref{fig:KrylovQFT}, we have plotted the Krylov complexity for this free scalar field theory. The complexity oscillates with time $t$, since we are only considering simple harmonic oscillator. As a result, this system doesn't feature chaos which is expected. So far, we have computed Krylov complexity with the space representation framework of \cite{Jefferson:2017sdb}. We can also apply this framework for the momentum representation of \cite{Chapman_2018}. In \cite{Chapman_2018}, the complexity is computed from the gaussian reference state to the target state by applying two mode squeezing operators. In section \ref{sec:KeqV}, we have given the Krylov complexity for the two mode squeezing operators using which Krylov complexity for the momentum variable framework of \cite{Chapman_2018} can be computed too. 

\begin{figure*}[htb!]
	\centering
	\includegraphics[width= 0.65
	\textwidth]{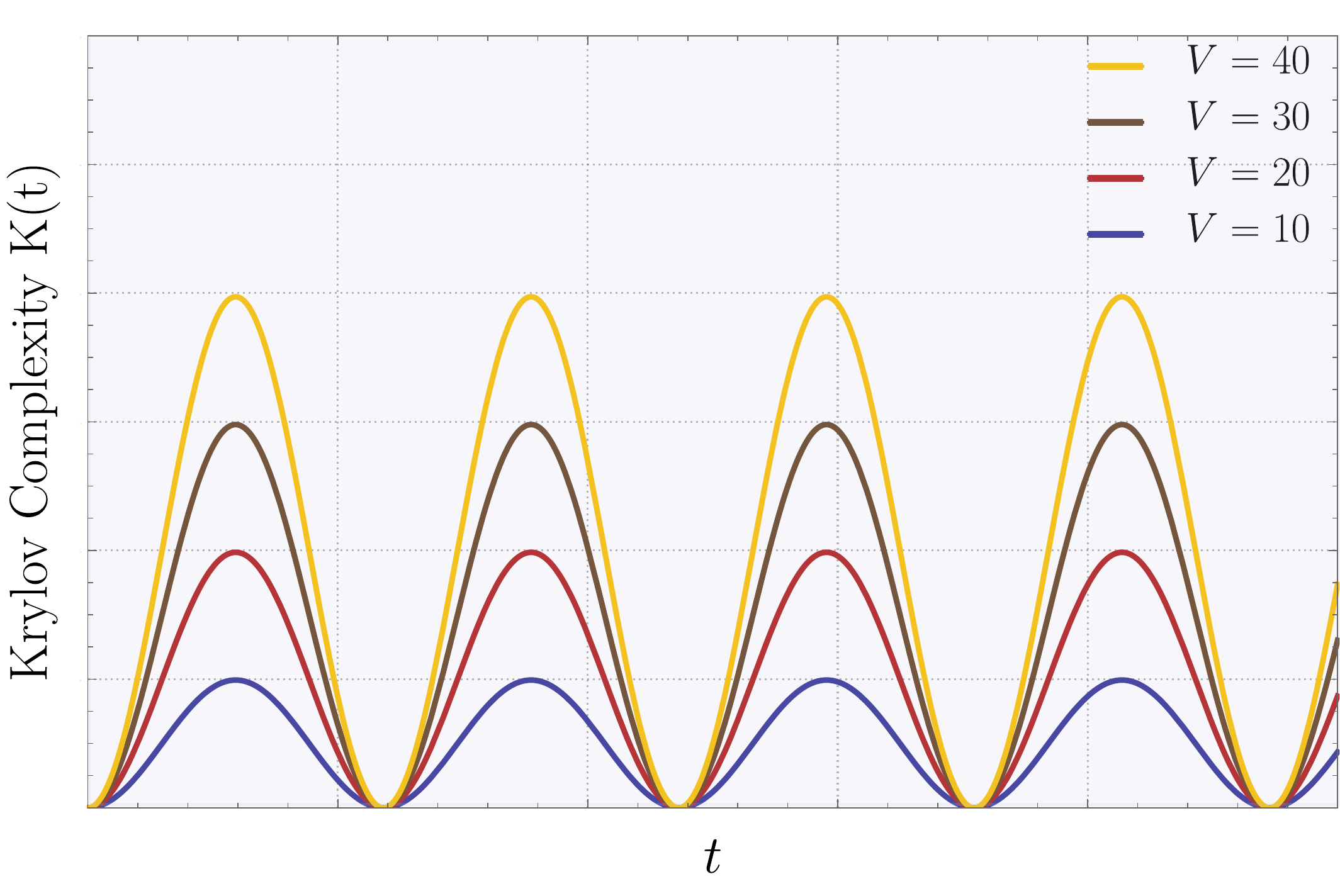}
	\caption{Behavior of Krylov complexity $K(t)$ as a function of time $t$ for different Volume in free scalar field theory.}
	\label{fig:KrylovHolo}
\end{figure*}

\section{{\textbf{ \large  Comparison with holography}}}
\label{sec:holography}
As we have mentioned before, the original motivation to study complexity in QFT comes from various holographic conjectures. So in this section, we will now compare the result for Krylov complexity obtained for quantum field theory with insights coming from holography. Even though, we are only studying for the free quantum field theory, the results obtained here actually satisfies some surprising qualitative checks.

In AdS/CFT conjectures, the properties of the bulk should be captured by the information in boundary. It turns out that the tool we often rely on which is entanglement entropy is not always sufficient for this task. Susskind et al \cite{Susskind:2014moa, Brown:2015lvg} proposed complexity to be such a possible candidate which can probe the physics behind the black hole horizons via “complexity = volume” and various other versions of it. The “complexity = volume” conjecture can be qualitatively expressed as:
\begin{equation}
    C_{\text{holo}} \approx \frac{V}{\delta^{d-1}}
\end{equation}
In the free scalar field theory we considered, the linear size of each spatial direction is $L = N\delta$, so the total (spatial) volume of the system is $V = L^{d-1} = N^{d-1}\delta^{d-1}$. The number of oscillators in terms of volume can be written as
\begin{equation}
N^{d-1} = \frac{V}{\delta^{d-1}}
\end{equation}
In this sense, if we can relate the complexity to be number of oscillators, then the complexity would then match with volume law similar to the holography conjectures. As we discussed on section \ref{sec:KeqV}, for the system like harmonic oscillator which is governed by $SL(2,R)$, Krylov complexity matches with the average particle number. Then in this picture, Krylov complexity equals exactly to the volume. This idea can be further explored for the thermofield double (TFD) state too which is dual to the eternal AdS black hole. In this framework, the boundary are entangled state of the two copies of the CFT which is connected by a wormhole, bulk. A perturbed TFD state can be thought of as a two-mode representation. As we showed in \ref{sec:KeqV}, for two-mode squeezed states, the Krylov complexity equals average particle number $\sinh^2r$. This further gives evidences for our case. One way to interpret this can be that for new space being emerged in the bulk, more and more degrees of freedoms need to be created in the boundary to effectively store the information. Then Krylov complexity could be related to this birth of new degrees of freedom. 

Finally, we can also make a comparison with the complexity obtained via Nielsen's method in \cite{Jefferson:2017sdb}
\begin{equation}
\label{eq:Nielsen}
C_{\text{Nielsen}}  \approx \left( \frac{V}{\delta^{d-1}} \right) \log\left( \frac{1}{\delta \omega_0} \right)
\end{equation}
In order to make the comparison, we have to make some assumptions. In QFT, the results are usually dominated by the UV modes, i.e, modes with $\tilde{\omega}_{\vec{k}} \approx 1/\delta$. Then the expression for Krylov complexity (\ref{eq:KrylovQFT}) becomes
\begin{equation}
    K(t) = N^{d-1} \frac{\sin^2( t/2\delta)}{\sinh^2(\beta/ 2 \delta)} = \frac{V}{\delta^{d-1}}\frac{\sin^2( t/2\delta)}{\sinh^2(\beta/ 2 \delta)} 
\end{equation}

In figure \ref{fig:KrylovHolo}, we have shown the behavior of Krylov complexity as a function of time for different volumes. As expected, complexity grows with volume. The change in complexity, when the spatial volume is fixed, is because of the time dependence of operator growth. The periodic nature of the complexity is due to considering free field theory. 
During early times $t$ and under further assumptions,
\begin{equation}
\label{eq:Kholo}
    K(t) = \frac{Vt}{\delta^{d-1}\beta} 
\end{equation}
Now, comparing with (\ref{eq:Nielsen}), it is clear that both Krylov and Nielsen's complexity grows with volume. However, Nielsen's measure (\ref{eq:Nielsen}) has some ambiguities, in particular the choice of $\omega_0$. Nielsen's complexity calculates the complexity from the reference to the target states, under some choice of gates and cost function. The $\omega_0$ is a parameter which characterizes the choice of the reference state. Therefore, this ambiguity in choice of reference state is now then reflected in the expression of Nielsen's complexity. There is also ambiguity in choice of gates and cost function. If we had chosen some other set of gates than in (\ref{eq:Nielsen}), then we would obtain other values of complexity though it is thought that the difference would not be huge. But, the choice of cost function strongly influences the expression of (\ref{eq:Nielsen}), in particular using $F_2$ cost function we would obtain the expression to be
\begin{equation}
    C_{\text{Nielsen}}  \approx \left( \frac{V}{\delta^{d-1}} \right)^{1/2} \log\left( \frac{1}{\delta \omega_0} \right)
\end{equation}
While these ambiguities can be avoided by choosing appropriate parameters, Krylov complexity has the benefit of being free from it. Instead Krylov complexity captures the operator spreading in time. Indeed during early times, Krylov complexity (\ref{eq:Kholo}) grows linearly with time.

\section{\large Towards interacting quantum field
theory}\label{sec:Desitter}
Keeping AdS/CFT correspondence in mind, it would be interesting to study interacting field theory. So far, we have focused on free scalar field theory. Though examining interacting QFT is beyond the scope of this paper, we will study field theory in which the inverted oscillator appears naturally. This theory exhibits chaotic features because of the inverted oscillator part. We will consider a theory with two free scalar fields deformed by a marginal coupling. In literature, this is also called (1+1)-dimensional $c=1$ conformal field theory. The procedure for computing complexity is similar to in section
 \ref{sec:qft}.  The Hamiltonian for this problem is the follow:

 \begin{equation}
\begin{aligned}
\label{eq:arpanHami}
   \text{H}&=\frac{1}{2}\int dx \Big[\pi_{1}^2+ (\partial_{x}\phi_1)^2+\pi_{2}^2 
   +(\partial_{x}\phi_2)^2+m^2 (\phi_1^2+\phi_2^2)\Big]\\
   &~~~~~~~~~~~~~~+ \lambda\int dx(\partial_{x}\phi_1)(\partial_{x}\phi_2)
\end{aligned}
\end{equation}

After regulating the theory, following procedure in \ref{qftlattice}, by placing it on a lattice with $\delta$ spacing with $\hat{m} = \frac{m}{\delta}, \bar{\omega} = \frac{1}{\delta^2}, \hat{\lambda} = \lambda \delta^{-4}$ and going to the normal coordinates $p_{a,k}, p_{b,k}, x_{a,k}, x_{b,k}$, we get the hamiltonian
\begin{align}
\label{eq:interact}
H=\frac{\delta}{2}\sum_{k=0}^{N-1}&\Big[p_{a,k}^2+\bar \omega_k^2  x_{a,k}^2+p_{b,k}^2+\omega_k^2 x_{b,k}^2\Big], 
\end{align}
where $\bar \omega_k^2 = \left(\hat m^2+4\,(\omega^2+\hat \lambda)\,\sin^2\Big(\frac{\pi\,k}{N}\Big)\right) $ and $ \omega_k^2 = \left(\hat m^2+4\,(\omega^2-\hat \lambda)\,\sin^2\Big(\frac{\pi\,k}{N}\Big)\right)$. For $\hat \lambda = 0$, this is just the expression for two free scalar field theory which we studied in section \ref{sec:qft}. Changing  $\hat \lambda$, $\omega_k$ can become negative  which is then described as a coupled inverted oscillators. However, $\bar \omega_k$ will always still stay positive, therefore it is always a regular oscillator.Then, we can write (\ref{eq:interact}) as a sum of regular as well as inverted oscillator for which the Krylov complexity using the examples in section \ref{sec:KeqV} is
\begin{equation}
    K(t) = \sum_{k=0}^{N-1} \left( \frac{\sin^2(\bar \omega_k t/2)}{\sinh^2(\beta \bar \omega_k /2)} + \frac{\sinh^2( \omega_k t/2)}{\sin^2(\beta  \omega_k /2)} \right)
\end{equation}
In figure \ref{fig:KrylovQFT}, we have plotted the behavior of Krylov complexity for both free and chaotic field theory. Because of the contributions coming from inverted oscillator, the complexity grows exponentially in time. This exponential growth of Krylov complexity is often associated with the the chaotic property. This makes sense since inverted oscillator is often considered as an example where chaos is experienced.

\section{\textbf{ \Large Conclusion}}\label{sec:Conclusion}
In this article, we have explored the Krylov complexity $K(t)$ in quantum field theory and make comparison with results from holography, and find surprising similarities.  Unlike Nielsen's complexity, Krylov complexity is free from ambiguities like choice of reference and target states, gate sets, and tolerances making it an ideal candidate to study in QFT and holography settings. Furthermore we showed that Krylov complexity, just like Nielsen's measure, also satisfies desirable properties like triangle inequality and continuity. A primary motivation to study complexity in QFT is to relate it with holographic conjectures like "Complexity = Volume". Here, we showed that when Krylov basis matches with Fock basis, the Krylov complexity equals average particle number showing that Krylov complexity scales with volume. Using the similar framework as for \cite{Jefferson:2017sdb, Chapman_2018}, we then compute the Krylov complexity for free scalar field theory, and find similarities with holography. We also extend this framework for field theory where inverted oscillator appears naturally and explore it's chaotic behavior.   

While we have explored the relationship of Krylov complexity with volume, we have not provided a rigorous proof of it. Using the insight of this article, one can try to find a proof of it. Since generalized displacement operator appeared multiple times in this article, studying it's particle probability distribution and relating with Krylov complexity would be a direction towards it. It seems like Krylov complexity is telling us about the particle production in the theory, it's relevance with holography also can be explored. Furthermore, one can also extend this formalism to interacting field theories, starting from simple $\phi^4$ theory and also in fermionic field theories. For most part of our work we have completely ignored quantum chaos. Since, Krylov complexity is often associated with chaos, exploring it in the context of quantum field theory seems to be important too. 

\textbf{Acknowledgement:}
~~~The Visiting Post Doctoral research fellowship of SC is supported by the J. C. Bose National Fellowship of Director, Professor Rajesh Gopakumar, ICTS, TIFR, Bengaluru. SC also would like to thank ICTS, TIFR, Bengaluru for providing the work friendly environment. SC also would like to thank the work friendly environment of The Thanu Padmanabhan Centre
For Cosmology and Science Popularization (CCSP), Shree Guru Gobind Singh Tricentenary (SGT) University, Gurugram, Delhi-NCR for providing tremendous support in research and offer the Assistant Professor (Senior Grade)
position.  SC also thank all the members of our newly formed virtual international non-profit consortium Quantum Structures of the Space-Time \& Matter (QASTM) for for elaborative discussions. KA and AR would like to thank the members of the QASTM Forum for useful discussions. Last but not least, we would like to acknowledge our debt to the people belonging to the various part of the
world for their generous and steady support for research
in natural sciences.


\begin{thebibliography}{99} 

\bibitem{Chapman:2021jbh}
S.~Chapman and G.~Policastro, ``{Quantum computational complexity from quantum
  information to black holes and back},''
  \href{http://dx.doi.org/10.1140/epjc/s10052-022-10037-1}{{\em Eur. Phys. J.
  C} {\bfseries 82} no.~2, (2022) 128},
  \href{http://arxiv.org/abs/2110.14672}{{\ttfamily arXiv:2110.14672
  [hep-th]}}.

\bibitem{Chapman:2021eyy}
S.~Chapman, D.~A. Galante, and E.~D. Kramer, ``{Holographic complexity and de
  Sitter space},'' \href{http://dx.doi.org/10.1007/JHEP02(2022)198}{{\em JHEP}
  {\bfseries 02} (2022) 198}, \href{http://arxiv.org/abs/2110.05522}{{\ttfamily
  arXiv:2110.05522 [hep-th]}}.

\bibitem{Faulkner:2022mlp}
T.~Faulkner, T.~Hartman, M.~Headrick, M.~Rangamani, and B.~Swingle, ``{Snowmass
  white paper: Quantum information in quantum field theory and quantum
  gravity},'' in {\em {2022 Snowmass Summer Study}}.
\newblock 3, 2022.
\newblock \href{http://arxiv.org/abs/2203.07117}{{\ttfamily arXiv:2203.07117
  [hep-th]}}.

\bibitem{Shaghoulian:2022fop}
E.~Shaghoulian and L.~Susskind, ``{Entanglement in De Sitter Space},''
  \href{http://arxiv.org/abs/2201.03603}{{\ttfamily arXiv:2201.03603
  [hep-th]}}.

\bibitem{Adhikari:2021ckk}
K.~Adhikari, S.~Choudhury, S.~Kumar, S.~Mandal, N.~Pandey, A.~Roy, S.~Sarkar,
  P.~Sarker, and S.~S. Shariff, ``{Circuit Complexity in $\mathcal{Z}_{2}$
  ${\cal EEFT}$},'' \href{http://arxiv.org/abs/2109.09759}{{\ttfamily
  arXiv:2109.09759 [hep-th]}}.

\bibitem{Brown:2017jil}
A.~R. Brown and L.~Susskind, ``{Second law of quantum complexity},''
  \href{http://dx.doi.org/10.1103/PhysRevD.97.086015}{{\em Phys. Rev. D}
  {\bfseries 97} no.~8, (2018) 086015},
  \href{http://arxiv.org/abs/1701.01107}{{\ttfamily arXiv:1701.01107
  [hep-th]}}.

\bibitem{Susskind:2014moa}
L.~Susskind, ``{Entanglement is not enough},''
  \href{http://dx.doi.org/10.1002/prop.201500095}{{\em Fortsch. Phys.}
  {\bfseries 64} (2016) 49--71},
  \href{http://arxiv.org/abs/1411.0690}{{\ttfamily arXiv:1411.0690 [hep-th]}}.

\bibitem{Brown:2015lvg}
A.~R. Brown, D.~A. Roberts, L.~Susskind, B.~Swingle, and Y.~Zhao,
  ``{Complexity, action, and black holes},''
  \href{http://dx.doi.org/10.1103/PhysRevD.93.086006}{{\em Phys. Rev. D}
  {\bfseries 93} no.~8, (2016) 086006},
  \href{http://arxiv.org/abs/1512.04993}{{\ttfamily arXiv:1512.04993
  [hep-th]}}.

\bibitem{Maldacena:2015waa}
J.~Maldacena, S.~H. Shenker, and D.~Stanford, ``{A bound on chaos},''
  \href{http://dx.doi.org/10.1007/JHEP08(2016)106}{{\em JHEP} {\bfseries 08}
  (2016) 106}, \href{http://arxiv.org/abs/1503.01409}{{\ttfamily
  arXiv:1503.01409 [hep-th]}}.

\bibitem{Nielsen1}
M.~A. Nielsen, ``A geometric approach to quantum circuit lower bounds,''.

\bibitem{Nielsen2}
M.~A. Nielsen, ``Quantum computation as geometry,''
  \href{http://dx.doi.org/10.1126/science.1121541}{{\em Science} {\bfseries
  311} no.~5764, (Feb, 2006) 1133–1135}.
  \url{http://dx.doi.org/10.1126/science.1121541}.

\bibitem{Nielsen3}
M.~R. Dowling and M.~A. Nielsen, ``The geometry of quantum computation,'' {\em
  Quantum Info. Comput.} {\bfseries 8} no.~10, (Nov., 2008) 861–899.

\bibitem{Nielsen4}
M.~A. Nielsen, M.~R. Dowling, M.~Gu, and A.~C. Doherty, ``Optimal control,
  geometry, and quantum computing,''
  \href{http://dx.doi.org/10.1103/PhysRevA.73.062323}{{\em Phys. Rev. A}
  {\bfseries 73} (Jun, 2006) 062323}.
  \url{https://link.aps.org/doi/10.1103/PhysRevA.73.062323}.

\bibitem{Balasubramanian:2022tpr}
V.~Balasubramanian, P.~Caputa, J.~Magan, and Q.~Wu, ``{A new measure of quantum
  state complexity},'' \href{http://arxiv.org/abs/2202.06957}{{\ttfamily
  arXiv:2202.06957 [hep-th]}}.

\bibitem{Caputa:2021sib}
P.~Caputa, J.~M. Magan, and D.~Patramanis, ``{Geometry of Krylov complexity},''
  \href{http://dx.doi.org/10.1103/PhysRevResearch.4.013041}{{\em Phys. Rev.
  Res.} {\bfseries 4} no.~1, (2022) 013041},
  \href{http://arxiv.org/abs/2109.03824}{{\ttfamily arXiv:2109.03824
  [hep-th]}}.

\bibitem{Parker:2018yvk}
D.~E. Parker, X.~Cao, A.~Avdoshkin, T.~Scaffidi, and E.~Altman, ``{A Universal
  Operator Growth Hypothesis},''
  \href{http://dx.doi.org/10.1103/PhysRevX.9.041017}{{\em Phys. Rev. X}
  {\bfseries 9} no.~4, (2019) 041017},
  \href{http://arxiv.org/abs/1812.08657}{{\ttfamily arXiv:1812.08657
  [cond-mat.stat-mech]}}.

\bibitem{Roberts:2018mnp}
D.~A. Roberts, D.~Stanford, and A.~Streicher, ``{Operator growth in the SYK
  model},'' \href{http://dx.doi.org/10.1007/JHEP06(2018)122}{{\em JHEP}
  {\bfseries 06} (2018) 122}, \href{http://arxiv.org/abs/1802.02633}{{\ttfamily
  arXiv:1802.02633 [hep-th]}}.

\bibitem{Rabinovici:2020ryf}
E.~Rabinovici, A.~S\'anchez-Garrido, R.~Shir, and J.~Sonner, ``{Operator
  complexity: a journey to the edge of Krylov space},''
  \href{http://dx.doi.org/10.1007/JHEP06(2021)062}{{\em JHEP} {\bfseries 06}
  (2021) 062}, \href{http://arxiv.org/abs/2009.01862}{{\ttfamily
  arXiv:2009.01862 [hep-th]}}.

\bibitem{Barbon:2019wsy}
J.~L.~F. Barb\'on, E.~Rabinovici, R.~Shir, and R.~Sinha, ``{On The Evolution Of
  Operator Complexity Beyond Scrambling},''
  \href{http://dx.doi.org/10.1007/JHEP10(2019)264}{{\em JHEP} {\bfseries 10}
  (2019) 264}, \href{http://arxiv.org/abs/1907.05393}{{\ttfamily
  arXiv:1907.05393 [hep-th]}}.

\bibitem{Jian:2020qpp}
S.-K. Jian, B.~Swingle, and Z.-Y. Xian, ``{Complexity growth of operators in
  the SYK model and in JT gravity},''
  \href{http://dx.doi.org/10.1007/JHEP03(2021)014}{{\em JHEP} {\bfseries 03}
  (2021) 014}, \href{http://arxiv.org/abs/2008.12274}{{\ttfamily
  arXiv:2008.12274 [hep-th]}}.

\bibitem{Dymarsky:2019elm}
A.~Dymarsky and A.~Gorsky, ``{Quantum chaos as delocalization in Krylov
  space},'' \href{http://dx.doi.org/10.1103/PhysRevB.102.085137}{{\em Phys.
  Rev. B} {\bfseries 102} no.~8, (2020) 085137},
  \href{http://arxiv.org/abs/1912.12227}{{\ttfamily arXiv:1912.12227
  [cond-mat.stat-mech]}}.

\bibitem{Dymarsky:2021bjq}
A.~Dymarsky and M.~Smolkin, ``{Krylov complexity in conformal field theory},''
  \href{http://dx.doi.org/10.1103/PhysRevD.104.L081702}{{\em Phys. Rev. D}
  {\bfseries 104} no.~8, (2021) L081702},
  \href{http://arxiv.org/abs/2104.09514}{{\ttfamily arXiv:2104.09514
  [hep-th]}}.

\bibitem{Hornedal:2022pkc}
N.~H\"ornedal, N.~Carabba, A.~S. Matsoukas-Roubeas, and A.~del Campo,
  ``{Ultimate Physical Limits to the Growth of Operator Complexity},''
  \href{http://arxiv.org/abs/2202.05006}{{\ttfamily arXiv:2202.05006
  [quant-ph]}}.

\bibitem{Caputa:2021ori}
P.~Caputa and S.~Datta, ``{Operator growth in 2d CFT},''
  \href{http://dx.doi.org/10.1007/JHEP12(2021)188}{{\em JHEP} {\bfseries 12}
  (2021) 188}, \href{http://arxiv.org/abs/2110.10519}{{\ttfamily
  arXiv:2110.10519 [hep-th]}}.

\bibitem{Balasubramanian:2021mxo}
V.~Balasubramanian, M.~DeCross, A.~Kar, Y.~C. Li, and O.~Parrikar,
  ``{Complexity growth in integrable and chaotic models},''
  \href{http://dx.doi.org/10.1007/JHEP07(2021)011}{{\em JHEP} {\bfseries 07}
  (2021) 011}, \href{http://arxiv.org/abs/2101.02209}{{\ttfamily
  arXiv:2101.02209 [hep-th]}}.

\bibitem{Balasubramanian:2019wgd}
V.~Balasubramanian, M.~Decross, A.~Kar, and O.~Parrikar, ``{Quantum Complexity
  of Time Evolution with Chaotic Hamiltonians},''
  \href{http://dx.doi.org/10.1007/JHEP01(2020)134}{{\em JHEP} {\bfseries 01}
  (2020) 134}, \href{http://arxiv.org/abs/1905.05765}{{\ttfamily
  arXiv:1905.05765 [hep-th]}}.

\bibitem{Choudhury:2021brg}
S.~Choudhury, A.~Mukherjee, N.~Pandey, and A.~Roy, ``{Causality Constraint on
  Circuit Complexity from ${\cal COSMOEFT}$},''
  \href{http://arxiv.org/abs/2111.11468}{{\ttfamily arXiv:2111.11468
  [hep-th]}}.

\bibitem{Adhikari:2021ked}
K.~Adhikari, S.~Choudhury, H.~N. Pandya, and R.~Srivastava, ``{PGW Circuit
  Complexity},'' \href{http://arxiv.org/abs/2108.10334}{{\ttfamily
  arXiv:2108.10334 [gr-qc]}}.

\bibitem{Adhikari:2021pvv}
K.~Adhikari, S.~Choudhury, S.~Chowdhury, K.~Shirish, and A.~Swain, ``{Circuit
  complexity as a novel probe of quantum entanglement: A study with black hole
  gas in arbitrary dimensions},''
  \href{http://dx.doi.org/10.1103/PhysRevD.104.065002}{{\em Phys. Rev. D}
  {\bfseries 104} no.~6, (2021) 065002},
  \href{http://arxiv.org/abs/2104.13940}{{\ttfamily arXiv:2104.13940
  [hep-th]}}.

\bibitem{Choudhury:2020hil}
S.~Choudhury, S.~Chowdhury, N.~Gupta, A.~Mishara, S.~P. Selvam, S.~Panda, G.~D.
  Pasquino, C.~Singha, and A.~Swain, ``{Circuit Complexity From Cosmological
  Islands},'' \href{http://dx.doi.org/10.3390/sym13071301}{{\em Symmetry}
  {\bfseries 13} (2021) 1301},
  \href{http://arxiv.org/abs/2012.10234}{{\ttfamily arXiv:2012.10234
  [hep-th]}}.

\bibitem{Bhargava:2020fhl}
P.~Bhargava, S.~Choudhury, S.~Chowdhury, A.~Mishara, S.~P. Selvam, S.~Panda,
  and G.~D. Pasquino, ``{Quantum aspects of chaos and complexity from bouncing
  cosmology: A study with two-mode single field squeezed state formalism},''
  \href{http://dx.doi.org/10.21468/SciPostPhysCore.4.4.026}{{\em SciPost Phys.
  Core} {\bfseries 4} (2021) 026},
  \href{http://arxiv.org/abs/2009.03893}{{\ttfamily arXiv:2009.03893
  [hep-th]}}.

\bibitem{Bhattacharyya:2020art}
A.~Bhattacharyya, W.~Chemissany, S.~S. Haque, J.~Murugan, and B.~Yan, ``{The
  Multi-faceted Inverted Harmonic Oscillator: Chaos and Complexity},''
  \href{http://dx.doi.org/10.21468/SciPostPhysCore.4.1.002}{{\em SciPost Phys.
  Core} {\bfseries 4} (2021) 002},
  \href{http://arxiv.org/abs/2007.01232}{{\ttfamily arXiv:2007.01232
  [hep-th]}}.

\bibitem{Adhikari:2022oxr}
K.~Adhikari and S.~Choudhury, ``{${\cal C}$osmological ${\cal K}$rylov ${\cal
  C}$omplexity},'' \href{http://arxiv.org/abs/2203.14330}{{\ttfamily
  arXiv:2203.14330 [hep-th]}}.

\bibitem{Jefferson:2017sdb}
R.~Jefferson and R.~C. Myers, ``{Circuit complexity in quantum field theory},''
  \href{http://dx.doi.org/10.1007/JHEP10(2017)107}{{\em JHEP} {\bfseries 10}
  (2017) 107}, \href{http://arxiv.org/abs/1707.08570}{{\ttfamily
  arXiv:1707.08570 [hep-th]}}.

\bibitem{Bhattacharyya:2020rpy}
A.~Bhattacharyya, S.~Das, S.~Shajidul~Haque, and B.~Underwood, ``{Cosmological
  Complexity},'' \href{http://dx.doi.org/10.1103/PhysRevD.101.106020}{{\em
  Phys. Rev. D} {\bfseries 101} no.~10, (2020) 106020},
  \href{http://arxiv.org/abs/2001.08664}{{\ttfamily arXiv:2001.08664
  [hep-th]}}.

\bibitem{Chapman_2018}
S.~Chapman, M.~P. Heller, H.~Marrochio, and F.~Pastawski, ``Toward a definition
  of complexity for quantum field theory states,''
  \href{http://dx.doi.org/10.1103/physrevlett.120.121602}{{\em Physical Review
  Letters} {\bfseries 120} no.~12, (Mar, 2018) }.
  \url{https://doi.org/10.1103%2Fphysrevlett.120.121602}.

\bibitem{Bhattacharjee:2022vlt}
B.~Bhattacharjee, X.~Cao, P.~Nandy, and T.~Pathak, ``{Krylov complexity in
  saddle-dominated scrambling},''
  \href{http://arxiv.org/abs/2203.03534}{{\ttfamily arXiv:2203.03534
  [quant-ph]}}.

\bibitem{Perelomov:1986tf}
A.~M. Perelomov, {\em {Generalized coherent states and their applications}}.
\newblock 1986.

\bibitem{Haegeman:2011uy}
J.~Haegeman, T.~J. Osborne, H.~Verschelde, and F.~Verstraete, ``{Entanglement
  Renormalization for Quantum Fields in Real Space},''
  \href{http://dx.doi.org/10.1103/PhysRevLett.110.100402}{{\em Phys. Rev.
  Lett.} {\bfseries 110} no.~10, (2013) 100402},
  \href{http://arxiv.org/abs/1102.5524}{{\ttfamily arXiv:1102.5524 [hep-th]}}.

\bibitem{Nozaki:2012zj}
M.~Nozaki, S.~Ryu, and T.~Takayanagi, ``{Holographic Geometry of Entanglement
  Renormalization in Quantum Field Theories},''
  \href{http://dx.doi.org/10.1007/JHEP10(2012)193}{{\em JHEP} {\bfseries 10}
  (2012) 193}, \href{http://arxiv.org/abs/1208.3469}{{\ttfamily arXiv:1208.3469
  [hep-th]}}.

\end{thebibliography}

\onecolumngrid 

\section{Appendix}

\subsection{Lancoz algorithm}\label{sec:lancoz}
The time evolution of  $\mathcal{W}(t) $ is
\begin{equation}
    \label{eq:timeSeries}
    \mathcal{W}(t) = e^{i\mathcal{L}t}\mathcal{W} = \sum_{n=0}^\infty \frac{(it)^n}{n!}\mathcal{L}^n \mathcal{W} = \sum_{n=0}^\infty \frac{(it)^n}{n!}\tilde{\mathcal{W}}_n
\end{equation}
Eq. (\ref{eq:timeSeries})  can be interpreted as a Schrödinger's equation where $\mathcal{W}(t)$ are "operator's wave functions", $\mathcal{L}$ is the Hamiltonian and "Hilbert space vectors" are
\begin{equation}
    \mathcal{W} \superequiv |\mathcal{\tilde{W}}), \mathcal{L}^1\mathcal{W} \superequiv |\mathcal{\tilde{W}}_1), \mathcal{L}^2\mathcal{W} \superequiv |\mathcal{\tilde{W}}_2),
    \mathcal{L}^3\mathcal{W} \superequiv |\mathcal{\tilde{W}}_3),\dots
\end{equation}
There is no  guarantee that $|\mathcal{\tilde{W}}_n)$ form an orthonormal basis a prior but with Lanczos algorithm, we can construct orthonormal Krylov basis 
$|\mathcal{W}_n)$. For this, we can choose Wightman norm. The algorithm is further simplified recognizing that first two operators are orthogonal on Wightman norm. Therefore, they form the Krylov basis
\begin{equation}
   |\mathcal{W}_0) :=   |\mathcal{\tilde{W}}_0) = |\mathcal{W}), |\mathcal{W}_1) := b_1^{-1} \mathcal{L} |\mathcal{\tilde{W}}_0)
\end{equation}
where $b_1 = \sqrt{(\mathcal{\tilde{W}}_0\mathcal{L}|\mathcal{L}\mathcal{\tilde{W}}_0)}$ normalized the vector. Iteratively, other states can be obtained as
\begin{equation}
    |A_n) = \mathcal{L}|\mathcal{W}_{n-1})-b_{n-1}|\mathcal{W}_{n-2})
\end{equation}
with normalization
\begin{equation}
    |\mathcal{W}_n) = b_n^{-1}|A_n), b_n = \sqrt{(A_n|A_n)}
\end{equation}
Then we continue lancoz algorithm continues till $b_n =0$. Together with Krylov basis, we can also obtain the lancoz coefficients $b_n$ which characterize the chaos of the system. 
Having done this, operator $\mathcal{W}(t)$ becomes
\begin{equation}
    |\mathcal{W}(t)) = e^{i\mathcal{L}t}|\mathcal{W}) = \sum_n i^n \phi_n(t)  |\mathcal{W}_n)
\end{equation}
where $\phi_n(t)$ are real amplitudes which can be obtained by solving the "Schrodinger equation" (\ref{eq:schrodinger}). The evolution equation is
\begin{equation}
    \partial_t \phi_{n}(t)  = b_n \phi _ {n-1} - b_{n+1}\phi _ {n+1}
\end{equation}
which can be solved with Lanczos coefficients $b_n$
and some initial condition. Finally we can give an expression for Krylov complexity/K-complexity 
\begin{equation}
    K = \sum_n n |\phi_n|^2
\end{equation}
One benefit of the Lancoz algorithm is that it can also characterize the chaotic property of the system. 
 Lancoz coefficients is conjectured to be bounded linearly \cite{Parker:2018yvk} as
\begin{equation}
    b_n \leq \epsilon n + \eta
\end{equation}
where $\epsilon$ and $\eta$ are obtained from the hamiltonian and gives information about its chaotic nature. Often, we can characterize the chaotic growth with Lyapunov exponent $\gamma$ as
\begin{equation}
    \gamma = 2 \epsilon
\end{equation}

\subsection{Lattice QFT}
\label{qftlattice}
After discretizing the free scalar theory (\ref{eq:roHami}) in square lattices, the theory becomes a collection of coupled harmonic oscillators
$H = \frac{1}{2} \sum_{\Vec{n}} H_1 + H_2$, where
\begin{equation}
   H_1 = \frac{p(\Vec{n})^2}{\delta^{d-1}} 
\end{equation}
and 
\begin{equation}
    H_2 = \delta^{d-1} \left[ \frac{1}{\delta^2} \sum_i  \left( \phi(\Vec{n}) - \phi(\Vec{n} - \hat{x_i})\right)^2 + m^2\phi(\Vec{n})^2   \right]
\end{equation}
where $\Vec{n}$ represents the position on lattice. Then, the theory becomes an large collections of coupled harmonic oscillators which is a well studied popular quantum mechanical problem.  As we mentioned before, in order to obtain the Krylov complexity for this complicated system, we can first obtain the complexity for even simpler system of two coupled harmonic oscillators and generalize it. The Hamiltonian of the two coupled Harmonic oscillator is given by:
\begin{equation}
H=\textstyle \frac{1}{2}\Big[p_{1}^{2}+p_{2}^{2}+\omega^{2}\left(x_{1}^{2}+x_{2}^{2}\right)+\Omega^{2}\left(x_{1}-x_{2}\right)^{2}\Big] .
\end{equation}
where $x_1$ and $x_2$ specify the spatial position, $\omega$ is the frequency of individual masses, and $\Omega$ is the inter-mass coupling. In order to solve this Hamiltonian, we can express the Hamiltonian in terms of normal coordinates $\bar{x}_{0}, \bar{x}_{1}, \tilde{p}_{0}, \tilde{p}_{1} $
\begin{equation}
\text{H}=\textstyle \frac{1}{2}\Big[  \tilde{p}_0^2 + \tilde{p}_1^2 + \tilde{\omega}_0^2\tilde{x}_0^2+ \tilde{\omega}_1^2\tilde{x}_1^2  \Big] .
\end{equation}
where, $\tilde{\omega}_{0}^{2}=\omega^{2}$ and $\tilde{\omega}_{1}^{2}=\omega^{2}+2 \Omega^{2}$.
Having done this, we can now generalize it to an arbitrary number of coupled oscillators. For simplicity, we will begin with $N$ oscillators on a one-dimensional circular lattice i.e. on $d= 1+1$ space-time dimension whose Hamiltonian is given by:
\begin{equation}
\label{eq: decopupledHO}
    H = \frac{1}{2} \sum_{a=0}^{N-1} \left[ p_a^2 + \omega^2x_a^2  + \omega^2(x_a- x_{a+1})^2 \right]
\end{equation}
where, periodic boundary condition implies $x_{a+N} =x_a $ and we have set $M_a = 1$. 
To solve this, we can go to Normal modes by using Discrete Fourier Transformation and then the Hamiltonian becomes
\begin{equation}
\label{eq:h2}
    H = \frac{1}{2} \sum_{k=0}^{N-1} \left[  |\tilde{p}_k|^2 + \tilde{\omega}_k^2 |\tilde{x}_k|^2 \right]
\end{equation}
where, $\tilde{\omega}_k^2 = \omega^2 + 4\Omega^2 \sin^2 \frac{\pi k }{N} $. 
We can follow the similar procedure for the theory (\ref{eq:arpanHami}).

\end{document}